\input harvmac

\noblackbox

\newcount\figno
\figno=0
\def\fig#1#2#3{
\par\begingroup\parindent=0pt\leftskip=1cm\rightskip=1cm\parindent=0pt
\global\advance\figno by 1
\midinsert
\epsfxsize=#3
\centerline{\epsfbox{#2}}
\vskip 12pt
{\bf Fig. \the\figno:} #1\par
\endinsert\endgroup\par
}
\def\figlabel#1{\xdef#1{\the\figno}}
\def\encadremath#1{\vbox{\hrule\hbox{\vrule\kern8pt\vbox{\kern8pt
\hbox{$\displaystyle #1$}\kern8pt}
\kern8pt\vrule}\hrule}}
\def\underarrow#1{\vbox{\ialign{##\crcr$\hfil\displaystyle
 {#1}\hfil$\crcr\noalign{\kern1pt\nointerlineskip}$\longrightarrow$\crcr}}}

\overfullrule=0pt

\newbox\tmpbox\setbox\tmpbox\hbox{\abstractfont}
\Title{\vbox{\baselineskip12pt\hbox{\hss hep-th/0410170}
\hbox{}}}
{\vbox{
\centerline{$U(n)$ Vector Bundles on Calabi-Yau Threefolds}
\bigskip
\centerline{for String Theory Compactifications}}}
\smallskip
\centerline{Bj\"orn Andreas}
\smallskip
\centerline{\it{Institut f\"ur Mathematik, Humboldt Universit\"at zu
Berlin}}
\centerline{\it{D-10115 Berlin, Germany}}
\medskip
\centerline{and}
\medskip
\centerline{Daniel Hern\'andez Ruip\'erez}
\smallskip
\centerline{\it{Departamento de Matem\'aticas, Universidad de
Salamanca}}
\centerline{\it{37008 Salamanca, Spain}}
\bigskip\bigskip

\noindent An explicit description of the spectral data of stable $U(n)$ vector bundles on elliptically fibered Calabi-Yau threefolds is given, extending previous work of Friedman, Morgan and Witten. The characteristic classes are computed and it is shown that part of the bundle cohomology vanishes. The stability and the dimension of the moduli space of the $U(n)$ bundles are discussed.  
As an application, it is shown that the $U(n)$ bundles are capable to solve the basic topological
constraints imposed by heterotic string theory. Various explicit solutions of the Donaldson-Uhlenbeck-Yau
equation are given. The heterotic anomaly cancellation condition is analyzed; as a result 
an integral change in the number of fiber wrapping five-branes is found. This gives a definite prediction for the number of three-branes in a dual $F$-theory model. The net-generation number
is evaluated, showing more flexibility compared with the $SU(n)$ case.

\Date{October, 2004}

\lref\NaSe{M. S. Narasimhan and C. S. Seshadri, ``Deformations of the Moduli Space
of Vector Bundles over an Algebraic Curve'', Ann. of Math. (2) {\bf 82} (1965) 540.}

\lref\Rama{ A. Ramanathan, ``Stable Principal Bundles on a Compact Riemann Surface,''
Math. Annalen 213 (1975) 129.}

\lref\SDon{ S. Donaldson, ``Antiselfdual Yang-Mills Connections on Complex Algebraic Surfaces
and Stable Bundles,'' Proc. London Math. Soc. {\bf 3} (1985) 1.}

\lref\AyBo{ M. F. Atiyah and R. Bott, ``The Yang-Mills Equations over Riemann Surfaces,''
Philos. Trans. Roy. Soc. London. Ser. A {\bf 308} (1982) 523."}

\lref\UhYa{ K. Uhlenbeck and S.-T. Yau, ``On the Existence of Hermitian Yang-Mills Connections
on Stable Bundles over Compact K\"ahler Manifolds,'' Comm. Pure App. Math. {\bf 39} (1986) 257,
{\bf 42} (1986) 703.}

\lref\Bor{ A. Borel, ``Sous-groupes Commutatifs et Torsion des Groupes de Lie Compacts Conexes,'' Tohoku Journal Ser. 2 (1961) {\bf 13} 216-240. }

\lref\AtiHir{ M. F. Atiyah and F Hirzebruch, ``Analytic Cycles on Complex Manifolds'', Topology {\bf 1} (1962) 25.}

\lref\Loo{E. Looijenga, ``Root systems and elliptic curves'', Invent. Math. {\bf 38} (1977) 17-32.}

\lref\Looj{E. Looijenga, ``Invariant theory for generalized root systems''.  Invent. Math.  {\bf 61}  (1980) 1-32}

\lref\ati{ M. F. Atiyah, ``Vector Bundles over an Elliptic Curve,'' Proc. London Math. Soc. {\bf 7}
(1957) 414-452.}

\lref\DoZh{M. R. Douglas and C. Zhou, {``Chirality Change in String Theory,''} hep-th/0403018.}

\lref\DoShZe{M. R. Douglas, B. Shiffman and S. Zelditch, {``Critical Points and Supersymmetric Vacua,''} math.cv/0402326.}

\lref\Do{M. R. Douglas, {``Statistics of String Vacua,''}, hep-ph/0401004.}

\lref\AsDo{S. Ashok and M. R. Douglas, {``Counting Flux Vacua,''} J. High Energy Phys. {\bf 0401} (2004) 060,
hep-th/0307049.}

\lref\DoSt{M. R. Douglas, {``The Statistics of String/ M-Theory Vacua,''}  J. High Energy Phys.  2003,  no. 5, 046, 61 pp. (electronic),
hep-th/0303194.}  

\lref\AC{B. Andreas and G. Curio, ``Three-Branes and Five-Branes in 
N=1 Dual String Pairs,'' Phys. Lett. {\bf B417} (1998) 41, hep-th/9706093.}

\lref\ACu{B. Andreas and G. Curio, ``On Discrete Twist and Four Flux in
N=1 Heterotic/F-Theory Compactifications'',  Adv. Theor. Math. Phys.  3  (1999),  no. 5, 1325, hep-th/9908193.}

\lref\ACC{B. Andreas and G. Curio,``Horizontal and Vertical Five-Branes in Heterotic/F-Theory Duality'',  JHEP {\bf 0001} (2000) 013, hep-th/9912025.}

\lref\Hart{R. Hartshorne, ``Algebraic Geometry'', Springer 1977.}

\lref\Mar{M. Maruyama,
{``On Boundedness of Families of Torsion Free Sheaves,''} J.Math.Kyoto Univ., {\bf 21} (1981), 673-701}

\lref\Tyau{ G. Tian and S.-T. Yau, ``Three Dimensional Algebraic manifolds with $c_1=0$ 
and $chi=-6$'', in Mathematical Aspects of String Theory, Proc. San Diego (1986) 543.}

\lref\CHS{ C. G. Callan, J. A. Harvey and A. Strominger, Nucl. Phys. {\bf B359} (1991) 611.}

\lref\CHSt{ C. G. Callan, J. A. Harvey and A. Strominger, Nucl. Phys. {\bf B367} (1991) 60.}

\lref\WIT{ E. Witten,  {``Small Instantons in String Theory''},  Nucl.
Phys. {\bf B460} (1996) 541.}

\lref\DMM {M.
Duff, R. Minasian and E. Witten,  {``Evidence for Heterotic/Heterotic
Duality''},  Nucl. Phys. {\bf B465} (1996) 413, { hep-th/9601036}.}

\lref\mv{D.R. Morrison and C. Vafa, {``Compactifications of F-Theory
on Calabi-Yau Threefolds II,''} Nucl. Phys. {\bf B476} (1996) 437-469.}

\lref\BDO{E. Buchbinder, R. Donagi and B. A. Ovrut, {``Vector Bundle
Moduli and Small Instanton Transitions''},  JHEP {\bf 0206} (2002) 054,
{\tt hep-th/0202084}.}

\lref\OUR{B. Andreas, G. Curio, D. Hern\'andez Ruip\'erez and S.T. Yau,
{``Fourier-Mukai Transform and Mirror Symmetry for D-Branes on
Elliptic Calabi-Yau''},
{ math.AG/0012196.}}

\lref\OURs{B. Andreas, G. Curio, D. Hern\'andez Ruip\'erez and S.T. Yau,
{``Fibrewise T-duality for D-branes on elliptic Calabi-Yau'', JHEP {\bf 0103} (2001), 020, hep-th0101129.}}

\lref\atmp{B. Andreas and D. Hern\'andez Ruip\'erez, {``Comments on $N=1$ Heterotic String Vacua''}, Avd. Theor. Math. Phys. {\bf 7} (2004) 751, hep-th/0305123.}

\lref\bart{C. Bartocci, U. Bruzzo, D. Hern\'andez Ruip\'erez and J.
M. Mu\~noz  Porras, {``Relatively stable bundles over elliptic
fibrations,''}  Mathematische Nachrichten {\bf 238} (2002) 23.}

\lref\RPo{D. Hern\'andez Ruip\'erez and J. M. Mu\~noz Porras,
{``Stable Sheaves on Elliptic Fibrations,''} J. Geom. Phys. {\bf 43}
(2002) 163.}

\lref\Simp{C. Simpson, {``Moduli of Representations of the
Fundamental Group of a Smooth Projective Variety. I,''} Publ. Math.
Inst. Hautes Etudes Sci. {\bf 79} (1994), 47-129.}

\lref\LuTe{M. L\"ubke and A. Teleman, {``The Kobayashi-Hitchin correspondence''} World Scientific, 1995.}

\lref\Kob{S. Kobayashi, {``Differential geometry of complex vector bundles''}, Publ. Math. Soc. Japan, {\bf 15}, Iwanami Shoten Pub. and Princeton Univ. Pres, 1987.}

\lref\Muk{S. Mukai, ``Duality between $D(X)$ and $D(\hat X)$ with
its application to Picard sheaves'',  Nagoya Math.~J.~{\bf 81} (1981),
153.}

\lref\Mukt{S. Mukai, ``Fourier functor and its application to the
moduli of bundles on an abelian variety'', Adv. Studies Pure Math. {\bf 10}
(1987), 515-550.}

\lref\OPP{B. A. Ovrut, T. Pantev and J. Park, {`` Small Instanton Transitions in Heterotic M-Theory,''}  JHEP {\bf 0005} (2000) 045, 
hep-th/0001133.}

\lref\Yo{K. Yoshioka, ``Moduli spaces of stable sheaves on
abelian surfaces'', Math. Ann. {\bf 312} (2001) 817-885.}

\lref\Ho{P. Horja, ``Derived category automorphisms from mirror symmetry'', Duke Math. J. (to appear), mathAG/0103231.}

\lref\hart{R. Hartshorne, ``Residues and Duality'', Springer LNM, {\bf 20} (1966).}

\lref\FMW{R. Friedman, J. Morgan and E. Witten, {``Vector Bundles and F-Theory,''}
Comm. Math. Phys. {\bf 187} (1997) 679, hep-th/9704151.}

\lref\BJPS{M. Bershadsky, A. Johansen, T. Pantev and V. Sadov, {``On
Four-Dimensional Compactifications of F-Theory''},  Nucl. Phys. {\bf
B505} (1997) 165-201, hep-th/9701165.}

\lref\FMWt{R. Friedman, J. Morgan and E. Witten, {``Vector Bundles over Elliptic Fibrations''}
J. Algebraic Geometry {\bf 8} (1999) 279.}

\lref\FMWthree{R. Friedman, J. W. Morgan and E. Witten, ``Principal G Bundles over Elliptic Curves'',
Math. Res. Lett. {\bf 5} (1998) 97-118, alg-geom/9709029.}

\lref\donaO{R. Donagi, ``Taniguchi Lectures on Principal Bundles on Elliptic Fibrations'', Integrable systems and algebraic geometry (Kobe/Kyoto, 1997),  33-46, World Sci. Publishing, River Edge, NJ, 1998, hep-th/9802094.}

\lref\AspDon{P.S. Aspinwall and R.Y. Donagi, ``The Heterotic String,
the Tangent Bundle, and Derived Categories'',
Adv. Theor. Math. Phys. {\bf 2} (1998) 1041,
hep-th/9806094.}

\lref\DLO{R. Donagi, A. Lukas, B.A. Ovrut and D. Waldram, {``Holomorphic
Vector Bundles and Non-Perturbative Vacua in M-Theory,''} JHEP {\bf
9906} (1999) 034, hep-th/9901009.}

\lref\DoOv{R. Donagi, B. A. Ovrut, T. Pantev and D. Waldram, {``Standard-Model Bundles''}, 
Adv. Theor. Math. Phys.{\bf 5} (2002) 563-615, math.ag/0008010.}

\lref\bridge{T. Bridgeland, `` Fourier-Mukai transforms for elliptic
surfaces'', J. Reine Angew. Math. {\bf 498}, (1998) 115-113, alg-geom/9705002.}

\lref\BridgeMac{T. Bridgeland and A. Maciocia, ``Fourier-Mukai
transforms for K3 and elliptic fibrations'', J. Algebraic Geometry {\bf 11} (2002), 629-657, 
math.AG/9908022.}

\lref\bbhm{C. Bartocci, U. Bruzzo, D. Hern\'andez Ruip\'erez and J.
M. Mu\~noz  Porras, ``Mirror symmetry on K3 surfaces via
Fourier-Mukai transform'', Commun. Math. Phys. {\bf 195} (1998), 79-93}

\lref\bbh{C. Bartocci, U. Bruzzo and D. Hern\'andez Ruip\'erez, ``A Fourier-Mukai transform for stable bundles on K3 surfaces'', J. Reine Angew. Math.. {\bf 496} (1997), 1-16}

\lref\Mac{A. Maciocia, ``Generalized Fourier-Mukai Transforms'',
J. reine angew. Math., {\bf 480} (1996) 197-211,
alg-geom/9705001.}

\lref\And {B. Andreas, {``On Vector Bundles and Chiral Matter in N=1
Heterotic String Compactifications''},  JHEP {\bf 9901} (1999) 011, hep-th/9802202.}

\lref\C{G. Curio,  {``Chiral Matter and Transitions in Heterotic
String Theory''},  Phys. Lett. {\bf B435} (1998) 39-48, {hep-th/9803224}.}

\lref\diac{D.-E. Diaconescu and G. Ionesei, {``Spectral Covers, Charged
Matter and Bundle Cohomology''},  JHEP {\bf 9812} (1998) 001,
{hep-th/9811129}.}

\lref\HuWi{C. Hull and E. Witten, {``Supersymmetric Sigma Models and the Heterotic String''},
Phys. Lett. {\bf 160B} (1985) 398.}

\lref\Witnew{E. Witten, {``New Issues in Manifolds of $SU(3)$
Holonomy''}, Nucl. Phys. {\bf B268} (1986) 79-112.}

\lref\DisGree{J. Distler and B. Greene, ``Aspects of (2,0) String Compactifications,'' Nucl. Phys. {\bf B304}
(1988) 1.}

\lref\CHSW{P. Candelas, G. Horowitz, A. Strominger and E. Witten, ``Vacuum Configurations for Superstrings'', Nucl. Phys. {\bf B258} (1985) 46-74.}

\lref\DisKa{J. Distler and S. Kachru, {``Duality of (0,2) String Vacua''},  Nucl. Phys. {\bf B 442}  (1995) 64, hep-th/9501111.}

\lref\HA{B. Andreas and D. Hern\'andez Ruip\'erez, in preparation.}

\lref\mdoug{M.R. Douglas, ``Topics in D-geometry. Strings '99 (Potsdam)'',  Classical Quantum Gravity  17  (2000),  no. 5, 1057, hep-th/9910170.}

\lref\doug{M.R. Douglas, ``D-Branes on Calabi-Yau Manifolds'',  European Congress of Mathematics, Vol. II (Barcelona, 2000),  449-466, Progr. Math, 202, BirkhŠuser, Basel, 2001, 
math.AG/0009209.}

\lref\Dou{M.R. Douglas, ``D-branes, categories and $N=1$ supersymmetry. Strings, branes, and M-theory''.  J. Math. Phys.  42  (2001),  no. 7, 2818-2843, hep-th/0011017.}

\lref\bdlr{I. Brunner, M. R. Douglas, A. Lawrence and C. R\"omelsberger,
{``D-Branes on the Quintic''},  JHEP {\bf 0010} (2000) 016, hep-th/9906200.}

\lref\diacrom{D.-E. Diaconescu and C. R\"omelsberger, ``D-Branes and Bundles
on Elliptic Fibrations'',  Nucl. Phys. {\bf B574}  (2000) 245, hep-th/9910172.}

\lref\kllw{P. Kaste, W. Lerche, C. A. L\"utken and J. Walcher,
``D-Branes on K3-Fibrations'',  Nucl. Phys. {\bf B582}  (2000) 203, hep-th/9912147.}

\lref\DouFioRoem{M.R. Douglas, B. Fiol and C. R\"omelsberger,
``The spectrum of BPS branes on a noncompact Calabi-Yau'',
hep-th/0003263.}

\lref\SMhM{R. Donagi, B. A. Ovrut, T. Pantev and D. Waldram, ``
Standard Models from Heterotic M-Theory'', hep-th/9912208.}

\lref\GH{P. Griffiths, J. Harris, ``Principles of Algebraic Geometry'', Wiley Interscience Publ., 1978.}

\newsec{Introduction}

Three approaches to construct holomorphic vector bundles, with structure group the complexification
$G_{\bf C}$ of a compact Lie group $G$, on elliptically fibered Calabi-Yau threefolds have been introduced in \FMW. The parabolic bundle approach applies for any simple $G$. One considers deformations of certain minimally unstable $G$-bundles corresponding to special maximal parabolic subgroups of $G$. The spectral cover approach applies for $SU(n)$ and $Sp(n)$ bundles and can be 
essentially understood as a relative Fourier-Mukai transformation. The del Pezzo surface approach applies for $E_6$, $E_7$ and $E_8$ bundles and uses the relation between subgroups of $G$ and
singularities of del Pezzo surfaces. Various aspects of these approaches have been further explored
in \refs{\BJPS\FMWt\FMWthree\donaO\AspDon\DLO\DoOv\bridge\BridgeMac\bbhm
\bbh\Mac\And\C-\diac}. 

The present paper continues the discussion of the spectral cover approach. Our aim is to give an explicit description of the spectral cover data and characteristic classes of stable $U(n)$ vector bundles $V$ on elliptically fibered Calabi-Yau threefolds $X$. As an application, we will study the question whether  the $U(n)$ bundles, which have non-zero first Chern class, are capable to solve the topological conditions
\eqn\topolc{c_1(V)=0 \ ({\rm mod}\ 2), \ \ \ ch_2(X)=ch_2(V),} 
that have to be imposed \refs{\HuWi,\Witnew,\DisGree} for a consistent heterotic string compactification. The first condition guarantees that the bundle $V$ admits spinors, while the second condition is the known anomaly cancellation condition which has to be generalized in the presence of five-branes. 

In addition, supersymmetry requires \CHSW, the connection on a given vector bundle has to satisfy the Donaldson-Uhlenbeck-Yau equation \UhYa, which can we written as ($J$ denotes the K\"ahler form of $X$)
\eqn\intya{\int_X c_1(V)\wedge J^2=0}
and implies that the vector bundle $V$ has to have zero-slope. Due to the zero-slope condition, the search for explicit solutions of \topolc\ has been restricted mainly to vector bundles with vanishing first Chern class (see for instance \refs{\DisGree,\DisKa}). If $c_1(V)$ is non-zero, than one typically has to work on manifolds for which the dimension of $H^{1,1}(X)$ is greater than or equal to two; a condition which is satisfied for the class of elliptically fibered Calabi-Yau threefolds considered in this paper. 

Let us also note, while the importance of vector bundles with vanishing first Chern class for the heterotic string theory is clear from a physics point of view, there is no reason to prefer bundles with vanishing first Chern class as long as mathematics is concerned. Actually, moduli spaces of stable or semistable vector bundles (or sheaves) with arbitrary Chern classes have been constructed. Also most mathematical operations with vector bundles and sheaves, for instance, twisting by a line bundle or elementary transforms, do not preserve the vanishing of the first Chern class. 

Clearly solving the topological constraints is only a first step to understand
heterotic string compactifications based on $U(n)$ vector bundles. Further issues such as the embedding properties of the $U(n)$ structure group into $E_8$ or the cancellation of an $U(1)$ anomaly arising on the string world sheet, have to be addressed \HA. 

\bigskip\noindent{\it Significance for $D$-Branes at Large Volume}

To conclude this introduction, let us mention a second application of $U(n)$ vector bundles (or more
generally sheaves) on Calabi-Yau manifolds. 

It has been shown that $D$-branes on Calabi-Yau manifolds, in the large volume limit, can be described by vector bundles (or sheaves) \refs{\bdlr\mdoug\doug-\Dou}. For $D$-branes wrapping all of the Calabi-Yau manifold, supersymmetry requires that the holomorphic connection $A$ on $V$ has to satisfy  
\eqn\duyz{F_A^{1,1}\wedge J^{2}={2\pi\over i}\mu(V)I_V.}

Since the $U(n)$ vector bundles (and sheaves) we consider in this paper are stable, they provide
solutions to \duyz, in contrast to heterotic string theory where one has to impose the stronger (zero-slope) condition $\mu(V)=0$. 

In particular vector bundles (or sheaves) with non-zero first Chern class
have been predicted, using the correspondence between the spectrum of $D$-branes at large volume and the Gepner point \bdlr, at which a particular conformal field theory is exactly solvable. The basic strategy which had been applied to establish this correspondence relied on a comparison of two central charges: the central charge corresponding to an integral vector in the middle cohomology lattice $H^3(Y,{\bf Z})$ of the mirror manifold $Y$ of a given Calabi-Yau threefold and the central charge associated to a $D$-brane state represented by a K-theory class and measured by the Mukai vector in the even cohomology. The comparison of these central charges gives a relation between the low energy charges and the topological invariants of the K-theory class and leads in many cases to a prediction of vector bundles (and sheaves) that have non-zero first Chern class. For elliptically and $K3$ fibered Calabi-Yau manifolds this correspondence has been worked out in many examples \refs{\diacrom\kllw\DouFioRoem-\OUR}. 

Thus finding stable vector bundles (or sheaves) with non-zero first Chern class, gives a direct proof that the corresponding moduli spaces, predicted in this correspondence, are not empty. 

Another advantage to treat vector bundles (or sheaves) as $D$-branes on elliptically 
fibered Calabi-Yau threefolds is that the relative Fourier-Mukai transformation can be 
viewed as a fiberwise T-duality transformation \refs{\OUR,\SMhM,\OURs}. This gives a concrete physical interpretation of the spectral cover approach for vector bundles.

This paper is organized as follows. In section 2, we first review the spectral cover approach 
as originally introduced in \FMW, then we review the more general description of vector bundles
and sheaves in terms of a relative Fourier-Mukai transformation. In section 3.1, we analyze the relation
of the Fourier-Mukai transformation and its inverse, this leads to a first simple 
example of an $U(n)$ vector bundle. It also shows that every stable $SU(n)$ vector bundle constructed 
in \FMW\ automatically determines a stable $U(n)$ vector bundle via the inverse Fourier-Mukai transformation. In section 3.2, we compute the characteristic classes of $U(n)$ vector bundles and comment on the finiteness of vector bundles (and sheaves) in this construction. In section 3.3, we derive the precise conditions which the spectral data of $U(n)$ vector bundle have to satisfy. In section 3.4, we study the reduction to $SU(n)$ bundles and recover the known expressions for the spectral data and characteristic classes derived in \FMW. In section 3.5, we review the known arguments and conditions for the stability of vector bundles which have been constructed in the spectral cover approach, or equivalently, by a relative Fourier-Mukai transformation. This section prepares also the analysis of the Donaldson-Uhlenbeck-Yau equation presented in section 4.1.
In section 3.6, we compute the bundle cohomology and show that the $U(n)$ bundles have no sections
which is relevant for the evaluation of the net-number of generations in a heterotic string compactification. 
In section 3.7, we discuss the moduli of $U(n)$ vector bundles. In section 4.1, we show that $U(n)$ bundles are capable to solve the Donaldson-Uhlenbeck-Yau equation, we give several explicit examples illustrating that. In section 4.2, we discuss the anomaly constraint and find a change in the number of five-branes wrapping the elliptic fiber. This leads to a definite prediction for the number of three-branes required for a consistent $F$-theory compactification. 
In section 4.3, we give the expression for the net-number of generations which receives a new term
compared to the $SU(n)$ case. 

There are three appendices. In appendix A, we provide some background on the Hermite-Einstein
equations. In appendix B, we review the Grothendieck-Riemann-Roch computation which led
to the characteristic classes of $U(n)$ vector bundles in section 3.2. In appendix C, we give a simplified derivation of the net-generation number for $SU(n)$ vector bundles which are constructed via the inverse Fourier-Mukai transformation. The same result has been obtained in \diac\ using Leray spectral sequence
technics.

\newsec{Review}

In this section we review the basic strategy of Friedman, Morgan and Witten \FMW\ which led to the construction of vector bundles on elliptically fibered Calabi-Yau threefolds $X$ with a section $\sigma$.
In the following we will denote by $E$ a single elliptic curve, i.e., a two-torus with a complex structure
and a distinguished point $p$, the identity element in the group law on $E$. We are interested in $G$-bundles on $E$ and $X$. We also recall the coherent description of vector bundles (and sheaves)
in terms of a relative Fourier-Mukai transformation which will be our essential tool in this paper.

\bigskip\noindent{\it $SU(n)$ Bundles on $E$}\smallskip

A $SU(n)$ bundle on $E$ is a rank $n$ vector bundle $V$ of trivial determinant.
Any $SU(n)$ vector bundle $V$ on $E$ can be expressed, as a smooth bundle,
in the form $V=\oplus_{i=1}^n{\cal L}_i$ where
${\cal L}_i$ are holomorphic line bundles\foot{In fact Atiyah showed for a smooth vector bundle $V$ on $E$ and $G=GL_r({\bf C})$ that that $V=\oplus_{i=1}^nL_i$ as smooth bundles, where each ${\cal L}_i$ is indecomposable and of degree zero \ati.}. Note that, as a holomorphic vector bundle, $V$ has only a filtration
with quotients given by the ${\cal L}_i$'s. The $SU(n)$ condition is imposed if $\otimes_{i=1}^n{\cal L}_i$ is the trivial line bundle. Further, $V$ is semistable if all ${\cal L}_i$ are of degree zero. The ${\cal L}_i$ 
are uniquely determined up to permutations and further determine
a unique point $q_i$ in $E$; conversely, every $q_i$ in $E$ determines a degree zero line
bundle ${\cal L}_i={\cal O}(q_i-p)$ on $E$. Further, a semistable smooth $SU(n)$ bundle $V=\oplus_{i=1}^n{\cal L}_i$ is determined due to the additional condition $\sum_{i=1}^n(q_i-p)=0$ (as divisors). 
The $q_i$ can be realized as roots of 
\eqn\szo{s=a_0+a_2x+a_3y+a_4x^2+a_5x^2y+\ldots+a_nx^{n/2}=0}
and give a moduli space of bundles ${\bf P}^{n-1}$, as stated by a theorem of Looijenga \refs{\Loo,\Looj}.
Note that if $n$ is odd, the last term in $s=0$ is given by $a_nx^{(n-3)/2}y$. The roots are determined by the coefficients $a_i$ only up to an overall scale factor so that the $a_i$ become the homogeneous coordinates on ${\bf P}^{n-1}$. 

\bigskip\noindent{\it $SU(n)$ Bundles on $X$}\smallskip

The basic strategy of Friedman, Morgan and Witten \FMW\ to construct $SU(n)$ vector bundles on $X$ is a two-step process. First, they use the bundle description of $SU(n)$ bundles on $E$ and then ``glue'' the bundle data together over the base manifold $B$ of $X$. More precisely, the variation of the $n$ points in the fiber over $B$ leads to a hypersurface
$C$ embedded in $X$, that is, $C$ is a ramified $n$-fold cover -the spectral cover- of the base. 
The line bundle on $X$ determined by $C$ is given by
${\cal O}_X(C)={\cal O}_X(n\sigma)\otimes {\cal M}$
where $\sigma$ denotes the section and ${\cal M}$ is a line bundle on $X$ whose restriction to the fiber is of degree zero. It follows that the cohomology class of $C$ in $H^2(X,{\bf Z})$ is given by 
$[C]=n\sigma +\eta,$
where $\eta=c_1({\cal M})$. 
Let $\pi_C:C\rightarrow B$ and denote by $E_b$ the general elliptic fiber over a point $b$ in $B$, then we have $C\cap E_b=\pi_C^{-1}(b)=q_1+\ldots+q_n$ and $\sigma\cap E_b=p$. Now, each $q_i$ determines a line bundle ${\cal L}_i$ of degree zero on $E_b$ whose sections are the meromorphic functions on $E_b$ with first order poles at $q_i$ and vanishing at $p$. The restriction of $V$ to $E_b$ is then $V|_{E_b}=\oplus_{i=1}^n {\cal L}_i$ as smooth bundles. As $b$ moves in the base the ${\cal L}_i$ move in one to one correspondence with the $n$ points $q_i$ above $b$. This specifies a unique line bundle ${L}$ on $C$ such that $\pi_{C*}{L}=V|_B$. 
Thus in addition to $C$ one has to specify a line bundle ${L}$ on $C$ to completely specify a rank $n$ vector bundle on $X$. 

This procedure leads typically to $U(n)$ vector bundles and will be our starting point. 
In \FMW\ the authors are interested in $SU(n)$ vector bundles. Therefore they impose  
two further conditions: ${\cal M}$ if restricted to the fiber is the trivial bundle and ${L}$ has to be specified such that $c_1(V)=0$. For the study of $U(n)$ vector bundles we shall keep the first condition.

\bigskip\noindent{\it Coherent Description via Fourier-Mukai Transform}\smallskip

Fourier-Mukai transformations have been introduced originally in \Muk\ as a tool for studying coherent sheaves on abelian varieties (or complex tori) and the corresponding derived categories. The transform has been extended later to other classes of manifolds, the first example being a Fourier-Mukai transformation for certain K3 surfaces \bbh. Also the relative situation for families of abelian varieties has been analyzed in \Mukt. After that the Fourier-Mukai transformation has been studied in many references, both in the absolute and the relative cases \refs {\bridge,\BridgeMac,\bbhm,\Mac,\bart,\RPo,\Yo} and in special connection with mirror symmetry and D-branes \refs{\AspDon,\OUR,\OURs,\atmp} or heterotic/M-theory \refs{\DLO,\DoOv}. 

Now a coherent description of $V$ can be given using a relative Fourier-Mukai transform \refs{\AspDon,\bridge,\bbhm,\OUR,\OURs}. Note that any sheaf constructed as a relative Fourier-Mukai transformation in terms of spectral data has trivial restriction to the fibers, so that it may be considered as a parametrization of $SU(n)$ bundles. Moreover, sheaves on an elliptic fibrations having degree zero on the fibers, admit a spectral cover description and can be reconstructed by the inverse Fourier-Mukai transformation from their spectral data. This is the reason why the condition of vanishing degree on the fibers is usually imposed, an assumption we will adopt also in this paper. 

For the description of the Fourier-Mukai transform it is appropriate to work on
$X\times_B {\tilde X}$ where $\tilde X$ is the compactified relative
Jacobian of $X$. $\tilde X$ parametrizes torsion-free rank 1 and
degree zero sheaves on the fibers of $X\to B$  and it is actually
isomorphic with $X$ so that we can identify
$\tilde X$ with $X$. We have a diagram:
\eqn\mats{\matrix{\;\;\; X\times_B{X}&\buildrel \pi_2
\over\longrightarrow&{\;\; X}\cr\scriptstyle{\pi_1} \biggr \downarrow &
&\scriptstyle{\pi}\biggr \downarrow \cr     \;\;\; X&\buildrel
{\pi} \over\longrightarrow&\;\;B}}
and the {Poincar\'e\/} sheaf
\eqn\poincdef{
{\cal P}={\cal O}(\Delta)\otimes {\cal O}(- \pi_1^*\sigma)\otimes {\cal
O}(-\pi_2^*\sigma)\otimes q^*K_B^{-1}}
normalized to make ${\cal P}$ trivial along $\sigma \times \tilde{X}$and
$X\times \sigma$. Here $q=\pi_1\circ \pi_1=\pi_2\circ \pi_2$ and ${\cal
O}(\Delta)$ is the dual of the ideal sheaf of the diagonal, which is
torsion-free of rank 1.

We define two Fourier-Mukai transforms on the derived category $D(X)$ of bounded complexes of coherent sheaves on $X$ in the usual way
\eqn\FMdef{\eqalign{{\Phi}({{\cal G}})&=R\pi_{1*}(\pi_2^*({\cal G})\otimes {\cal P})\,,
\cr
\hat {\Phi}({\cal G})&=R\pi_{1*}(\pi_2^*({\cal G})\otimes
\hat{\cal P})\,.
}}
Here 
\eqn\poincinvdef{
\hat{\cal P}={\cal P}^*\otimes q^*K_B^{-1}}

We can also define the Fourier-Mukai functors ${\Phi}^i$ and
${\hat {\Phi} }^i$,
$i=0,1$ in terms of single sheaves $V$ by taking ${\Phi} ^i({V})$
and $\hat {\Phi}^i({V})$
as the $i$-th cohomology sheaves of the complexes ${\Phi}({V})$ and $\hat
{\Phi} ({V})$, we have
\eqn\lis{\eqalign{
{\Phi}^i({V})&=R^i\pi_{1*}(\pi_2^*({V})\otimes {\cal P})\,,
\cr
\hat {\Phi}  ^i({V})&=R^i\pi_{1*}(\pi_2^*({V})\otimes
\hat{\cal P})\,.}}

As it is usual, we say that a sheaf is $WIT_j$ with respect to $\Phi$ (or $\hat\Phi$) if
$\Phi^j({\cal F})=0$ (or $\hat\Phi^j({\cal F})=0$) for $j\neq i$. 

Further, we note that one obtains for any complex
${\cal G}$ (or any element in the category) 
an invertibility result
\eqn\rela{\Phi(\hat\Phi({\cal G}))={\cal G}[-1], \ \ \ \ \hat{\Phi}(\Phi({\cal G}))={\cal G}[-1].}
Concerning the meaning of the -1
shift, we first recall that if we denote by ${\cal G}^i$ the cohomology
sheaves of a complex ${\cal G}$, the cohomology sheaves of the shifted complex
${\cal G}[n]$ are  ${\cal G}[n]^i={\cal G}^{i+n}$; then if we
consider a complex given by a single sheaf
$V$ located at the ``degree zero'' position.
$\hat{\Phi}({\Phi}(V))= V[-1]$ means that the complex $\hat{\Phi}({\Phi}(V))$
has only one cohomology sheaf, that is $V$, but located at ``degree
1'', $[\hat{\Phi}({\Phi}(V))]^1=V,\ \ [\hat{\Phi}({\Phi}(V))]^i=0, i\neq 1$.
When $\Phi^0(V)=0$ the
complex $\Phi(V)$ reduces to a single sheaf, that is
the unique Fourier-Mukai transform $\Phi^1(V)$, but located at ``degree 1'', that
is, $S\Phi(V)=\Phi^1(V)[-1]$ and the
complex $\hat{\Phi}(\Phi(V))=\hat{\Phi}(\Phi^1(V))[-1]$ has two cohomology
sheaves, one at degree 1, given by $\hat{\Phi}^0(\Phi^1(V))$, and
one at degree 2, given by $\hat{\Phi}^1(\Phi^1(V))$. So one has
$\hat{\Phi}^0(\Phi^1(V))=V, \ \ \hat{\Phi}^1(\Phi^1(V))=0$.

\newsec{$U(n)$ Vector Bundles on Elliptic Calabi-Yau Threefolds}

In what follows we require that our elliptically fibered Calabi-Yau threefold $\pi\colon X\to B$ 
has a section $\sigma$ (in addition to the smoothness
of $B$ and $X$). This (and the Calabi-Yau condition) restricts the base $B$ to be a Hirzebruch surface
$({ F}_m$, $m\geq 0)$, a del Pezzo surface $(dP_k, k=0,...,8)$, a rational elliptic surface $(dP_9)$, blown-up Hirzebruch surfaces or an Enriques surface \refs{\DLO,\mv}. 

From \lis\ it follows that we can construct vector bundles on elliptically fibered Calabi-Yau 
threefolds $X$ using either a relative Fourier-Mukai transform or its inverse. In both cases one starts
from the spectral data $(C,L)$, where $C$ is a surface in $X$ and $L$ is a torsion free rank one sheaf on $C$ \foot{For reducible $C$, a sheaf is torsion free if it is pure of dimension one. A pure sheaf of dimension $i$ is a sheaf whose support has dimension $i$ and it has no
subsheaves concentrated on smaller dimension.}.
We take $C=n\sigma+\pi^*\eta$, more precisely $C\in |n\sigma+\pi^*\eta|$ with $\eta$ some effective curve in $B$. So we have two possibilities to define a sheaf 
on $X$
\eqn\vt{V=\Phi^0(i_*L), \ \ \  \pi_{C*}L=\sigma^*V}
\eqn\vo{W=\hat\Phi^0(i_*L), \ \ \  \pi_{C*}L=\sigma^*W\otimes K_B}
where $i\colon C\to X$ is the closed immersion of $C$ into $X$.
When $C$ is irreducible and $L$ is a line bundle, the first definition agrees with the description of vector bundles given by Friedman, Morgan and Witten which we reviewed above.

\subsec{Relation of $V$ to $W$}

We will now show that there is a relation between $V$ and $W$.
For this let us write $\tau\colon X\to X$ for the elliptic involution on
$X\to B$. We can consider the induced involution on $X\times_B X$ given by 
$\tau(x,y)=(\tau(x),y)=(-x,y)$
where $x$ and $y$ are points of $X$ such that $\pi(x)=\pi(y)$. One easily sees that $\pi_1\circ \tau= \tau\circ \pi_1$, $\tau\circ \pi_2=\pi_2$ and $\hat\tau^*{\cal P}={\cal P}^*$.
If we denote by $q=\pi_1\circ\pi=\pi_1\circ \pi$ the projection of $X\times_B X$ onto $B$, then for any object ${\cal G}$ in the derived category we have
\eqn\invtwo{\eqalign{ \hat\Phi ({\cal G})&= R\pi_{1*}(\pi_2^* {\cal G}\otimes {\cal P}^*\otimes q^*K_B^{-1}) = R\pi_{1*}(\tau^*(\pi_2^* {\cal G}\otimes {\cal P}))\otimes \pi^*K_B^{-1} \cr
&= \tau^* (R\pi_{1*}(\pi_2^* {\cal G}\otimes {\cal P}))\otimes \pi^*K_B^{-1} = \tau^*\Phi({\cal G})\otimes \pi^*K_B^{-1}
}}
by base change in the derived category \hart. Then in our current situation we get
\eqn\prerels{W=\tau^*V\otimes \pi^*K_B^{-1}.}
So if we start with a vector bundle $V$ (as given in \FMW) with vanishing first Chern class, the new bundle we obtain has non-zero first Chern class and gives our first simple example of an $U(n)$ vector bundle.

\subsec{Characteristic Classes of $V$ and $W$}

The goal in this section is to compute the characteristic classes of $V$ and $W$.

\smallskip\noindent{\it Computation of $ch(V)$}\smallskip

In order to do so, we first recall from Appendix B the formulae giving the topological invariants of both the Fourier-Mukai transform and the inverse Fourier-Mukai transform of a general complex $\cal G$ in the derived category. Using those formulae we find that, if we start with the sheaf $E=i_*L$  with Chern characters given by 
\eqn\chV{ch_0(i_*L)=0,\
ch_1(i_*L)=n\sigma+\pi^*{\eta},\
ch_2(i_*L)=\sigma \pi^*\eta_E+a_E F,\
ch_3(i_*L)={s_E}}
with $\eta_E, {\eta}\in H^2(B)$, then the Chern characters of the
Fourier-Mukai transform $V=\Phi^0(i_*L)$ of $i_*L$ are given by
\eqn\chVS{\eqalign{ch_0(V)&=n\cr
ch_1(V)&=\pi^*(\eta_E-{1\over 2}n c_1)\cr
ch_2(V)&=(-\pi^*{\eta})\sigma+({s_E} -{1\over 2}\pi^*\eta_E c_1\sigma
+{1\over12}n c_1^2\sigma)F\cr
ch_3(V)&=-a_E +{1\over 2}\sigma c_1
\pi^*{\eta}}}

\smallskip\noindent{\it Computation of $ch(W)$}\smallskip

The characteristic classes of $W$ can be determined directly from \prerels\ using two facts. First,  
the Chern character is multiplicative, thus applied to our situation we get $ch(W)=ch(\tau^*V)ch(\pi^*K_B^{-1})$; second, we note that all the sheaves $E$ we are considering (including $V$, $i_*L$ and $W$) have Chern characters of the type \chV.  All those classes are invariant under $\tau^*$ because both the fiber and the section are so; it follows that ${ch_i(\tau^*E)=ch_i(E)}$ for every $i\ge 0$. Thus we get $ch(W)=ch(V)ch(\pi^*K_B^{-1})$. Now because $B$ is a surface we have the relations $(\pi^*a)^3=\pi^*(a^3)=0$ and $(\pi^*\omega)\pi^*a=\pi^*(\omega a)=0$ where $a\in H^2(B)$ and $\omega\in H^4(B)$, using these we find
\eqn\chVSS{\eqalign{ch_0(W)&=n \cr
ch_1(W)&=\pi^*(\eta_E+{1\over 2}n c_1)\cr
ch_2(W)&=(-\pi^*{\eta})\sigma+({s_E} +{1\over 2}\pi^*\eta_E c_1\sigma
+{1\over12}n c_1^2\sigma)F\cr
ch_3(W)&=-a_E-{1\over 2}\sigma c_1
\pi^*{\eta}+n\sigma c_1^2}}
The above formulae follow also directly from appendix B taking into account that $W=\hat\Phi^0(i_*L)$ (see \vo). Note that the number of moduli of $W$ agrees with the number of moduli of the vector bundle $V$ since in general twisting a vector bundle by a line bundle $N$ does not change the dimension of the moduli space of $V$ as $h^1(End(V))=h^1(End(V\otimes N))$. 

\bigskip\noindent{\it Remark on Finiteness of Sheaves}

Having computed $ch(V)$ and $ch(W)$, one could ask 
whether there are bounds on possible values of $ch_i$. This type of question
fits into a recently initiated program which studies the vacuum selection problem of 
string/M-theory from a statistical point of view \refs{\DoZh\DoShZe\Do\AsDo-\DoSt}.
A partial answer can be given as a consequence of theorems by Maruyama \Mar\ which characterize the boundedness of families of torsion free sheaves. Maruyama proved (see Corollary 
4.9) that if we have a smooth projective complex manifold $X$ of dimension $r$ with a 
polarization $H$ and fix $n\in {\bf Z}$,  $c_1$ the numerical class of 
a divisor on $X$ and $a\in {\bf Z}$, then the set
$C_i(n,c_1,a)$ of the algebraic equivalence classes of $i$-th Chern 
classes $c_i(V)$ of all semistable vector bundles $E$ on $X$ with 
$c_1(V)=c_1$, $rk(V)=n$ and $c_2(V) H^{r-2} \le a$ is finite for all 
$2\le i\le r$. In particular, if $X$ is a three-dimensional Calabi-Yau manifold, and we fix  
$n$, $c_1$ and $c_2$, then the number of possible values of $c_3(V)$ 
for a semistable vector bundle $V$ of rank $n$ on $X$ with $c_1(V)=c_1$ and 
$c_2(V)=c_2$ is finite.

\subsec{Fixing $c_1$ of the Spectral Line Bundle}

One should note that $\eta_E$, $a_E$ and $s_E$ are not completely arbitrary. One needs to
ensure that there exists a line bundle $L$ on $C$ whose Chern characters are given by \chV\
which we will determine now. We also given an explicit expression for $c_1(L)$. 

For this we analyze the Grothendieck-Riemann-Roch theorem applied
to the $n$-sheeted cover $\pi_C\colon C\rightarrow B$ which gives
\eqn\Grr{ch(\pi_{C*}L)Td(B)=\pi_{C*}(ch(L)Td(C))}
from which we derive 
\eqn\ett{c_1(\sigma^*V)+{n\over 2}c_1(B)=\pi_{C*}\big(c_1(L)+{c_1(C)\over 2}\big).}
For $(1,1)$ classes $\alpha$ on $B$ we have $\pi_{C*}\pi_C^*\alpha=n\alpha$ and
$\sigma^*$ applied to $V$ gives $c_1(\sigma^*V)=\eta_E-{1\over 2}nc_1(B)$
so we get
\eqn\pIs{\pi_{C*}(c_1(L))=\pi_{C*}(-{c_1(C)\over 2}+{{\pi^*_C\eta_E}\over n})} 
which gives
\eqn\fch{c_1(L)=-{c_1(C)\over 2}+{{\pi^*_C\eta_E}\over n}+\gamma,}
where $\gamma\in H^{1,1}(C,{\bf Z})$ is some cohomology class satisfying $\pi_{C*}\gamma=0\in H^{1,1}(B,{\bf Z})$. The general solution for $\gamma$ has been derived in \FMW\ and is given by
$\gamma=\lambda(n\sigma_{\vert_{C}}-\pi_C^*\eta+n\pi_C^*c_1(B))$ with $\lambda$ some rational number which we will specify below. Let us also note $\gamma$ restricted to $S=C\cap \sigma$
is given by $\gamma_{\vert_{S}}=-\lambda\pi^*\eta(\pi^*\eta-n\pi^*c_1(B))\sigma$.

\bigskip\noindent{\it Specification of $\eta_E$}\smallskip

So far we assumed that $\eta_E$ is an element of $H^2(B)$. We need to find now an $\eta_E$ such that
$c_1(V)$ and $c_1(L)$ are both integer classes; otherwise we had to impose directly conditions on $n$ or $c_1(B)$. It turns out that there are many solutions for $\eta_E$ satisfying the integer class condition.
For our purposes it will be sufficient to restrict to the case $\eta_E={n\over 2}\beta$ with $\beta\in H^2(B,{\bf Z})$. This is in agreement with $ch_2(i_*L)$ being an element of $H^4(X,{\bf Q})$ and 
if we reduce to the $SU(n)$ case by setting $\eta_E={1\over2}nc_1(B)$, we find that \fch\ reduces to the expression for $c_1(L)$ of $SU(n)$ bundles derived in \FMW. 

For $\eta_E={1\over 2}n\beta$ we find that $c_1(V)$ is always an integer class if $n$ is even. If $n$ is odd
we have to impose the additional constraint $\beta=c_1(B)$ (mod 2). 

The conditions for the right hand side of \fch\ to be an integer class are given for $U(n)$ bundles with 
$n$ odd by $\lambda=m+{1\over 2}$ and $\beta=nc_1(B) \ ({\rm mod}\ 2)$ (with $m\in {\bf Z}$). 
Thus if $n$ is odd we have to find simultaneously solutions to $\beta=c_1(B) \ ({\rm mod}\ 2)$
and $\beta=nc_1(B) \ ({\rm mod}\ 2)$. This can be solved since $(n-1)c_1(B)=0 \ ({\rm mod}\ 2)$
and $(n-1)$ is even.

For $U(n)$ bundles with $n$ even we find two solutions, we can either work with $\lambda=m$ and $\eta=\beta \ ({\rm mod}\ 2)$ or with $\lambda=m+{1\over 2}$ and $\beta=nc_1(B) \ ({\rm mod}\ 2)$.

Let us briefly discuss how the conditions change if we modify $\eta_E$ slightly. For $\eta_E=n\beta$ we find again $c_1(V)$ is always an integer class if $n$ is even. However, if $n$ is odd
we find $c_1(B)=0 \ ({\rm mod}\ 2)$ which restricts the choice of $B$. There are two solutions for $n$ 
even such that $c_1(L)$ is an integer class: $\lambda=m$ and $\eta=0 \ ({\rm mod}\ 2)$ or $\lambda=m+{1\over 2}$. If $n$ is odd we find $\lambda=m+{1\over 2}$ and $c_1(B)=0 \ ({\rm mod}\ 2)$.

\bigskip\noindent{\it Specification of $a_E$ and $s_E$}\smallskip

Having fixed $c_1(L)$ we can now go on and determine $a_E$ and $s_E$ in terms of 
$\eta_E$. For this apply the Grothendieck-Riemann-Roch theorem to $i\colon C\rightarrow X$ 
which gives ${ch(i_*L)Td(X)=i_*(ch(L)Td(C))}.$
We note that $i_*(1)=C$ and using the fact that $i_*(c_1(B)\gamma)=0$ a simple computation 
gives 
\eqn\blu{\eqalign{a_E&=\gamma_{\vert_{S}}+{1\over n}\eta_E\eta\cr
s_E&={1\over 24}{nc_1^2}+{1\over 2n}{\eta_E^2}-\omega}}
where $\omega$ is given by
\eqn\pso{\omega=-{1\over 24}c_1(B)^2(n^3-n)+{1\over 2}\big(\lambda^2-{1\over 4}\big)n\eta(\eta-nc_1(B))}

\subsec{Reduction to $SU(n)$}

To make contact with the work of \FMW\ let us describe the reduction from 
$U(n)$ to $SU(n)$ and thus recover the second Chern class of an $SU(n)$ vector bundles
originally computed in \FMW\ and the third Chern class evaluated  in \refs{\And,\C,\diac}.

In order to describe the reduction to $SU(n)$ we specify
the class $\eta_E={1\over 2}nc_1$ giving $c_1(V)=0$. If we insert this into \blu\ we find the new expressions for $a_E$ and $s_E$ which are given by 
\eqn\red{\eqalign{
a_E&=\gamma_{\vert_{S}}+{1\over 2}c_1\eta\cr
s_E&={1\over 6}n c_1^2\sigma-\omega,}}
and with \chVS\ we find the Chern-classes of an $SU(n)$ vector bundle $V$ on the elliptic fibered Calabi-Yau threefold 
\eqn\chsdt{r(V)= n, \ \ \ c_1(V)= 0, \ \ \ c_2(V)= \pi^*(\eta)\sigma+\pi^*(\omega), \ \ \ c_3(V)=-2\gamma_{\vert_{S}}}
in agreement with \refs{\FMW,\And,\C,\diac}.

As we were enforcing $c_1(V)=0$ by setting $\eta_E={1\over 2}nc_1$, one could ask are there other possibilities to reduce $U(n)$ to $SU(n)$ vector bundles on $X$ such that the Chern classes differ from \chsdt. For this recall that it is often possible to twist a given vector bundle with non-zero first Chern class such that the resulting bundle has zero first Chern class. Assume we have $V$ with $c_1(V)\neq 0$. If we twist as $W=V\otimes N$  where $N$ is a line
bundle on $X$, we have $c_1(W)=c_1(V)+nc_1(N)$ (with $n$ the rank of $V$). Then $c_1(W)=0$ implies that $c_1(V)=-nc_1(N)$. A natural choice for $N$ would be to take a root of the inverse of the determinant line bundle, i.e., $c_1(N)=-{1/n}c_1(V)=c_1(detV^{-1/n}))$, assuming that it exists. 

Since all our bundles have vanishing first Chern class on the fibers, this implies that the restrictions of
$N$ to the fibers have also vanishing first Chern class, so they are trivial (the general fiber being an elliptic curve). Thus $N=\pi^* L$ for some line bundle $L$ on $B$.

Using $c_1(V)=\pi^*(\eta_E-{1\over 2}n c_1(B))$, the equation $c_1(V)+nc_1(N)=0$ reduces to
$\pi^*(nc_1(L)+\eta_E-{1\over 2}n c_1(B))=0$ and using $\eta_E={1\over 2}n\beta$ we get $c_1(L)+{1\over 2}\beta-{1\over 2} c_1(B)=0$. This
equation can be solved for $L$ if ${1\over 2} (\beta-c_1(B))$ is an integer class, and if so,
the space of solutions  is isomorphic to the Jacobian (or Picard scheme) of $B$.
Thus if $c_1(L)$ is represented by an integer class, all twists lead to $SU(n)$ vector bundles with characteristic classes given by \chsdt. 

\subsec{Stability of $V$ and $W$}

The discussion of stability of $V$ and $W$ depends on the properties of the defining data $C$ and $L$.
If $C$ is irreducible and $L$ a line bundle over $C$ then $V$ and $W$ will be vector bundles
stable with respect to
\eqn\pol{J=\epsilon J_0+\pi^*H_B, \qquad {\epsilon} > 0}
if $\epsilon$ is sufficiently small (cf. Theorem 7.1 in \FMWt\ where the statement is proven under the additional assumption that the restriction of $V$ to the generic fiber is regular and semistable). Here $J_0$ refers to some arbitrary K\"ahler class
on $X$ and $H_B$ a K\"{a}hler class on the base $B$. It implies that the bundle $V$ can be taken 
to be stable with respect to $J$ while keeping the volume
of the fiber $F$ of $X$ arbitrarily small compared to the volumes of effective curves
associated with the base. That $J$ is actually a good polarization can be seen by assuming 
$\epsilon=0$. Now one observes that  ${\pi}^*H_B$ is not a K\"{a}hler class on $X$ since
its integral is non-negative on each effective curve $C$ in $X$, however, there is one curve, the fiber $F$, where the integral vanishes. This means that ${\pi}^*H_B$ is on the boundary of
the K\"{a}hler cone and to make $V$ stable, one has to move slightly into the interior of the K\"{a}hler cone, that is, into the chamber which is closest to the boundary point ${\pi}^*H_B$. 

Also we note that although ${\pi}^*H_B$ is in the boundary of the K\"ahler cone, we can still define the slope $\mu_{{\pi}^*H_B}(V)$ with respect to it. Since $({\pi}^*H_B)^2$
is some positive multiple of the class of the fiber $F$, semi-stability with respect to ${\pi}^*H_B$ is implied by semi-stability of the restrictions $V {\vert}_F$ to the fibers.
Assume that $V$ is not stable with respect to $J$, then there is a
destabilizing sub-bundle $V' \subset V$ with $\mu_J(V') \ge \mu_J(V)$.
But semi-stability along the fibers says that $\mu_{{\pi}^*H_B}(V') \le
\mu_{{\pi}^*H_B}(V)$. If we had equality, it would follow that $V'$
arises by the spectral construction from a proper sub-variety of the
spectral cover of $V$, contradicting the assumption that this cover is
irreducible. So we must have a strict inequality $\mu_{{\pi}^*H_B}(V')
<\mu_{{\pi}^*H_B}(V)$. Now taking $\epsilon$ small enough, we can
also ensure that $\mu_{J}(V') < \mu_{J}(V)$ thus $V'$ cannot destabilize $V$.

\bigskip\noindent{\it Generalization to Reducible Spectral Covers}\smallskip

Let us now consider the case that $C$ is flat over $B$. If $C$ is not irreducible than there may exist line bundles such that $V=\Phi^0(i_*L)$ is not stable with respect to the polarization given by \pol, however,
the condition one has to impose to the spectral data in order that $V$ is a stable sheaf on $X$ with respect to \pol, has been derived in \atmp. Actually, if $C$ is flat over $B$ and $L$ is a pure dimension
sheaf on $C$ than $V$ is stable with respect to $\bar{J}=\bar\epsilon \sigma+\pi^* H_B$ for sufficiently small $\bar\epsilon$ if and only if $i_*L$ is stable with respect to this polarization. Let us note here that
stability with respect to \pol\ for $\epsilon$ sufficiently small is equivalent to stability with respect to $\bar J$ for sufficiently small $\bar\epsilon$ if we take $J_0=a\sigma+b\pi^*H_B$ for some positive $a$, $b$. 
Furthermore, note that if $C$ is irreducible and $L$ is a line bundle the latter condition is automatically satisfied. 

Moreover, $V$ and $W$ are simultaneously stable with respect to $\tilde J$. This is not a surprise because from $ch_i(\tau^* V)=ch(V)$ we know that $V$ and $\tau^*V$ are simultaneously stable and from $W=\tau^*V\otimes \pi^*K_B^{-1}$ we know that stability is the same for $W$ and $\tau^*V$.

Note that assuming $C$ is flat and $L$ a pure dimension one sheaf, one finds a larger class of stable sheaves $V$ (and $W$) than originally constructed in \FMW, because we do not need an irreducible spectral cover and that the restriction of the bundle (sheaf) to the generic fiber is regular. 

\subsec{Bundle Cohomology and Index Computation}

In this section we will show that the $U(n)$ bundles constructed via a relative Fourier-Mukai transform
have no sections. This simplifies the index of the Dirac operator with values
in the respective vector bundle which is related to the net-number of generations
of chiral fermions in heterotic string compactifications.

To begin let us recall, for stable $SU(n)$ vector bundles the Riemann-Roch theorem reduces to ${h^2(X,V)-h^1(X,V)={1\over 2}c_3(V)}$ since $h^0(X,V)$ and by Serre duality $h^3(X,V)$ vanish. Otherwise a nonzero element of $H^0(X,V)$ would define a mapping from the trivial line bundle into $V$ but since the trivial line bundle has rank one, this would violate the slope inequality. Since $V^*$ is stable as well,  Serre duality gives $h^3(X,V)=h^0(X,V^*)=0$. 

In case of vector bundles $E$ with arbitrary first Chern class the index 
is given by 
\eqn\inds{\sum_{i=0}^3(-1)^i{\rm dim}H^i(X,E)=\int ch_3(E)+{1\over 12}c_2(X)c_1(E).}

Let us now examine the left hand side of \inds\ and show that $h^0(X,E)=h^3(X,E)=0$ for 
$E$ given by a relative Fourier-Mukai transformation. That is, we assume that the restriction of $E$ to the fibers are semistable of degree zero. We also assume that the rank of $E$ is bigger than one. Then the spectral cover of $E$ is different from the section $\sigma$ because if it were $\sigma$, then $\Phi^1(E)=\sigma_*(N)$ for $N$ a line bundle on $B$ and one would find that $E=\pi^*(N\otimes K_B^{-1})$.

Since both $E$ and  the structure ${\cal O}_X$ sheaf are WIT$_1$ with respect to $\Phi$ with transforms  $\Phi^1(E)=i_*L$, $\Phi^1({\cal O}_X)=\sigma_* K_B$ where $\sigma\colon B\to X$ is the
section, we can apply the Parseval theorem for the relative Fourier-Mukai transform.

Assume that we have sheaves ${F}$, $\bar {G}$ that are
respectively WIT$_h$ and WIT$_j$ for certain $h$, $j$. The Parseval theorem says that one has
\eqn\preparc{Ext_X^i({F},{G})=Ext_X^{h-j+i}({\Phi}^h({F}),{\Phi}^j({G}))\,,}
thus giving a correspondence between the extensions of ${F}$,
${G}$ and the extensions of their Fourier-Mukai transforms.
The proof is very simple, and relays on two facts. The first one is
that the ext-groups can be computed in terms of the derived category, namely
\eqn\extdef{ Ext_X^i(F,G)=Hom_{D(X)}(F,G[i])}
The second one is that the Fourier-Mukai transforms of
${F}$, ${G}$ in the derived category $D(X)$ are
${\Phi}  ({F})={\Phi}  ^h({F})[-h]$,
${\Phi}  ({G})={\Phi}  ^j({G})[-j]$.
Now, since the Fourier-Mukai transform is an equivalence of categories, 
one has
\eqn\extder{
\eqalign{
Hom_{D(X)}({F},{G}[i]) & =Hom_{D(X)}({\Phi}  ({
F}),{\Phi}  (
{G}[i])) \cr
&=Hom_{D(X)}({\Phi}  ^h({F})[-h],{\Phi}  ^j( {G})[-j+i]) \cr
&=Hom_{D(X)}({\Phi}  ^h({F}),{\Phi}  ^j( {G})[h-j+i])}}
so that \extdef\ gives the Parseval theorem \preparc.
Thus we obtain (cf., \OUR\ or \atmp)
\eqn\parc{\eqalign{ Ext_X^i(E,{\cal O}_X)& =Ext_X^i(i_*L,\sigma_*(K_B)) \cr
Ext_X^i({\cal O}_X,E)& =Ext_X^i(\sigma_*(K_B), i_*L)
\,.\quad i\geq 0}}
Then we have
\eqn\hnot{H^0(X,E)=Hom_X({\cal O}_X,E)= Hom_X(\sigma_*(K_B), i_*L)=Hom_C(i^*(\sigma_*K_B), L) =0}
because $i^*(\sigma_*K_B)$ is concentrated on $S=C\cap \sigma$ and $L$ is a line bundle on $C$. 
Analogously we get
\eqn\hthree{H^3(X,E)=Hom_X(E,{\cal O}_X)= Hom_X(i_*L,\sigma_*(K_B))=Hom_B(\sigma^*(i_*L),K_B)=0,}
because $\sigma^*(i_*L)$ is concentrated on $S$ and $K_B$ is a line bundle on $B$.

The same reasoning applies to $\hat\Phi$, so that both $V$ and $W$ constructed in \vt\ and \vo\ as well as $\tau^*V$ have vanishing groups $H^0$ and $H^3$.

\subsec{Moduli of $U(n)$ Vector Bundles}

In this section let us briefly discuss whether the expression for the number of moduli of a rank $n$ vector bundle constructed in the spectral cover approach depends on the first Chern class or not. 
In case of a $X$ being a $K3$ surface it is known that the number of bundle moduli is given by
\eqn\kth{h^1(X, End(V))=2nc_2(V)-(n-1)c_1(V)^2-2n^2+2}
which follows from a Riemann-Roch index computation \Kob\ and shows the dependence 
on $c_1(V)$. This formula can be also derived using the spectral cover approach. For this 
let us recall that if $V$ is given by the spectral data $(C,L)$ we get \OUR\ or \OURs\
\eqn\din{h^1(X, End(V))=h^1({\cal O_C})+h^0(N_C)=C^2+2}
where $N_C$ denotes the normal bundle of the spectral cover in $X$ and the last equality 
follows from $h^0(N_C)=h^0(K_C)$ (because $X=K3$), which using Serre duality is equal to 
$h^1({\cal O}_C)$. A Grothendieck-Riemann-Roch index computation gives for $ch_1(i_*L)=C$ the following expression \RPo\
\eqn\chil{C=-c_1(V)+r\sigma-(ch_2(V)-c_1(V)\sigma)F}
and using \din\ we recover \kth.

Now in case of $X$ being a Calabi-Yau threefold the spectral cover $C$ does not depend 
on $c_1(V)$ so we expect the number of bundle moduli of a $U(n)$ vector bundle to agree
with the corresponding number of moduli of a $SU(n)$ vector bundle. This number is related to the number of parameters specifying the spectral cover $C$ and the dimension of the space of
holomorphic line bundles $L$ on $C$ determines the moduli. 
The first number is given by the dimension of the linear system
$|C|=|n\sigma+\eta|$. The second number is given by the dimension of
the Picard group $Pic(C)=H^1(C,{\cal O}^*_C)$ of $C$.
One thus expects the moduli of $V$ to be given by \BDO\
\eqn\dime{{\rm dim}H^1(X, {\rm End}(V))=\dim |C|+\dim Pic(C).}
This formula was proven in \OUR\ and \OURs\ under the equivalent form
\eqn\dintwo{h^1(X, End(V))=h^1({\cal O_C})+h^0(N_C)= h^{(0,1)}(C)+h^{(2,0)}(C)}
 when $C$ is a divisor in $X$ using the Parseval isomorphism 
\eqn\secondparc{Ext^1_X(V,V)=Ext^1_X(i_*L,i_*L)}
 and an argument of spectral sequences associated to Grothendieck duality for the closed immersion $i\colon C\to X$. Here \secondparc\ follows from  \preparc\ since $i_*L$ is WIT$_0$ with unique transform $V$.  

Still assuming that $C$ is a divisor, we have $N_C=K_C$ since $X$ is Calabi-Yau, and then $h^0(N_C)=h^2({\cal O_C})$, so that we get
\eqn\dinthree{h^1(X, End(V))=h^1({\cal O_C})+h^2({\cal O_C})}
(for $C$ connected). If we assume moreover that $C$ is smooth, then Lefschetz theorem on hyperplane sections (\GH, p. 156) implies that $h^1({\cal O_C})=0$ and then,
\eqn\dinfour{h^1(X, End(V))=h^2({\cal O_C})\,.}

\newsec{Application to Heterotic String Theory}

In this section we present a systematic analysis of the Donaldson-Uhlenbeck-Yau equation
and of the anomaly cancellation condition for $U(n)$ vector bundles on elliptic Calabi-Yau threefolds. We also give an explicit expression for the net-generation number $N_{gen}$ using the above index computation of section 3.6. 

\subsec{Solving the Donaldson-Uhlenbeck-Yau Equation}

As mentioned in the introduction, a geometric compactification of the ten-dimensional heterotic string is 
specified by a holomorphic stable $G$-bundle $V$ over a Calabi-Yau manifold $X$. The Calabi-Yau condition, the holomorphicity and stability of $V$ are a direct consequence of the required supersymmetry
in the uncompactified space-time. More precisely, supersymmetry requires that a connection $A$ on $V$ has to satisfy  
\eqn\hymz{{F_A^{2,0}=F_A^{0,2}=0, \ \ \ F_A^{1,1}\wedge J^{2}=0,}}
where the first condition implies the holomorphicity of $V$. A theorem by Uhlenbeck and Yau, which is a particular case of the Kobayashi-Hitchin correspondence, guarantees the existence and uniqueness of a solution to the second equation which can be written as (cf. Appendix A)
\eqn\duy{\int_X c_1(V)\wedge J^2=0.}
This equation can be solved if the first Chern class of a given vector bundle vanishes or if the underlying manifold $X$ has $h^{11}(X)$ greater or equal to 2; for $h^{11}(X)=1$ the first Chern class is a multiple of the K\"ahler form $J$ of $X$ and $J^3$ is never zero as its integral is the volume of $X$.

Let us discuss whether the stable $U(n)$ vector bundles with characteristic classes given by \chVS\
and \blu\ can solve the Donaldson-Uhlenbeck-Yau equation. The $U(n)$ bundles are stable with respect to $\bar{J}={\bar\epsilon}\sigma+\pi^*H_B$. To simplify the analysis, let us make the following assumptions: we take as before $\eta_E={1\over 2}n\beta$ with $\beta\in H^2(B,{\bf Z})$ and we set $H_B=c_1(B)$ which restricts us to work on base manifolds whose anti-canonical line bundle is ample, i.e., we consider del Pezzo surfaces which are isomorphic either to ${\bf P}^1\times {\bf P}^1$ or
${\bf P}^2$ with $k$ points ($0\leq k\leq8$) blown up, usually denoted by $dP_k$. Some of those surfaces
can be also viewed as Hirzebruch surfaces, ${\bf P}^1\times{\bf P}^1=F_0$ and $dP_1=F_1$. The Donaldson-Uhlenbeck-Yau equation can then be written as
\eqn\duyre{\pi^*(c_1(B)(\beta-c_1(B)))\sigma=0.}
Furthermore, we have to make sure that $c_1(V)$ and $c_1(L)$ are integer classes,
thus we have to solve in case $n$ is even equation \duyre\ simultaneously with $\beta=\eta \ ({\rm mod}\ 2)$ or $\beta=nc_1(B)\ ({\rm mod}\ 2)$ depending on whether we take $\lambda=m$ or $\lambda=m+{1\over 2}$. If $n$ is odd, we have to solve \duyre\ simultaneously with $\beta=c_1(B) \ ({\rm mod}\ 2)$ and $\beta=nc_1(B)\ ({\rm mod}\ 2)$.

To give a concrete example, let us analyze these conditions in more detail for the Hirzebruch surfaces $F_m$ with $m=0,1$ and an $U(n)$ bundle with $n$ being odd. Recall that $c_1(F_m)=2b+(2+m)f$ where $b^2=-m$, $bf=1$ and $f^2=0$ and since $F_m$ is a rational surface we have $c_1(F_m)^2=8$. Thus we have to look for solutions of (we set $c_1=c_1(B)$)
\eqn\esl{\beta c_1=8, \ \ \  \beta=c_1 \  ({\rm mod}\ 2), \ \ \  \beta=nc_1 \  ({\rm mod}\ 2).}
Since $\beta$ is a generic element of $H^2(F_m, {\bf Z})$ we take $\beta=xb+yf$ 
with $x$, $y$ being not restricted. Thus the first equation in \esl\ reduces to $x+y=4$. 
Now for $m=0$ we have $c_1=0 \ ({\rm mod}\ 2)$ thus the last two equations of \esl\
reduce to $\beta=0 \  ({\rm mod}\ 2 )$ and so we have to take $x=2x'$ and $y=4-2x'$ with $x' \in {\bf Z}.$ 

If $m=1$ the first equation in \esl\ reduces to $x+2y=8$. The last two equations in \esl\ 
can be written as $\beta=f \ ({\rm mod}\ 2)$ and $\beta=nf \ ({\rm mod}\ 2)$ which can be solved since
$(n-1)f=0 \ ({\rm mod}\ 2)$ and $n-1$ is even. Thus we only have to require $\beta=f \ ({\rm mod}\ 2)$, i.e.,
$xb+(y-1)f=0 \ ({\rm mod}\ 2)$ which restricts $x$ to be an even and $y$ to be odd number, i.e., we take
$x=2x'$ and $y=2y'+1$ with $x',y'\in{\bf Z}$, such that $x'+2y'=3$.

A similar analysis can be done for $U(n)$ bundles with $n$ even. For this we only have to note that
if we want the spectral surface $C$ to be irreducible we have to make sure that the linear system $|\eta|$  
is base point free in $B$ and $\eta-nc_1(B)$ is effective. For $B=F_m$ and $\eta=vb+wf$ the corresponding conditions have been derived in \OPP\ and are given by $v\geq 2m$, $w\geq n(m+2)$ and $w\geq m v$. Taking these into account one easily obtains solutions to $\beta c_1=8$ and $\eta=\beta \ ({\rm mod}\ 2).$

\bigskip\noindent{\it Solutions for Arbitrary Ample $H_B$}\smallskip

So far we assumed in our analysis that $H_B=c_1(B)$. Let us now study the situation
for arbitrary ample $H_B$ and $\eta_E={1\over 2}n\beta$ as before. In order to solve the Donaldson-Uhlenbeck-Yau equation, we must solve 
in addition to \duyre\ the constraint
\eqn\duyret{\pi^*(H_B(\beta-c_1(B)))\sigma=0.}

Now let us consider again $U(n)$ bundles with $n$ odd and take as base manifolds the Hirzebruch surfaces $F_m$ whose $c_1$ is even, the odd case will be discussed below. It is instructive to solve first \esl\ and then find solutions to \duyret.  Similar to our analysis above we start with a generic $\beta\in H^2(F_m,{\bf Z})$ and find solutions to \esl\ in terms of $x,y$ and $m$. The first equation in \esl\ reduces to $2x+2y-xm=8$. For $m$ even we have $c_1=0 \ ({\rm mod}\ 2)$ thus we find again the constraint $\beta=0 \ ({\rm mod}\ 2)$ which determines the coefficients of $\beta$ to be given by $x=2x'$ and $y=4+(m-2)x'$ and $x'\in {\bf Z}$. 

We can now solve \duyret. For this let us recall that he K\"ahler cone (the very ample classes) of $F_m$ is equals the positive (ample) classes and is given \Hart\ by the numerically effective classes $wb+zf=(w,z)$ with $w>0, z>mw$. So if we insert  $H_B=ub+vf$ and $\beta=(2x')b+(4+(m-2)x')f$ into \duyret, we find
the condition 
\eqn\vre{v=(1+{m\over 2})u,}
where $u>0$, so we have to restrict to the Hirzebruch surface $F_0$, otherwise $v>mu$ is violated. 

To complete the analysis let us consider $B=F_m$ with $m$ odd. From \esl\ we find
$\beta=mf \ ({\rm mod}\ 2)$ and $\beta=nmf \ ({\rm mod}\ 2)$ which can be solved since
$m(n-1)f=0 \ ({\rm mod}\ 2)$ is always satisfied. Thus to solve $\beta=mf \ ({\rm mod}\ 2)$
we have to take $\beta=xb+yf$ with $x=2x'$ and $y=2y'+1$ with $x',y'\in{\bf Z}$. Moreover, from $\beta c_1=8$ we find that $y'={1\over 2}(3-x'(2-m))$. If we insert these expressions in \duyret, we find
the condition (with $u>0$)
\eqn\vrell{v=(1+{x'm\over{2x'-2}})u,}
with $x'\neq 1$ and ${x'\over (2x'-2)}>{{m-1}\over m}$ which guarantees that $H_B$ is ample. Thus
if $x'>0 $ (and $m\neq 1$) we find $x'$ has to satisfy the bound $x'<{{2m-2}\over {m-2}}$ and if $x'<0$ (and $m\neq 1$) we find $x'>{{2m-2}\over {m-2}}$. For $m=1$ equation \vrell\ always satisfies $v>u$,
so there are no bounds for $x'$.

\subsec{The Anomaly Constraint}

In a heterotic string compactifications (without five-branes) on a Calabi-Yau threefold $X$ one has to specify, in addition to a stable holomorphic $G$ vector bundle $V$, a $B$-field on $X$ of field strength $H$. In order to get an anomaly free theory, the structure group $G$ of $V$ is fixed to be either $E_8\times E_8$ or $Spin(32)/{{\bf Z}_2}$ or one of their subgroups and $H$ has to satisfy the generalized Bianchi identity which, if integrated over a four-cycle in $X$, gives the topological constraint 
${c_2(X)=c_2(V).}$ This constraint has to be modified in the presence of five-branes and if $V$ has a non-zero first Chern class.

The inclusion of five-branes\foot{In \refs{\CHS, \CHSt} magnetic five-brane solutions to heterotic string theory were discussed, the gauge five-brane, the symmetric five-brane and the neutral five-brane. The gauge and symmetric five-brane solution involve finite size instantons of an unbroken
non-Abelian gauge group. Neutral five-branes can be interpreted as zero size instantons of the $SO(32)$ heterotic string \WIT.} changes this topological constraint \refs{\FMW,\DMM,\DisGree}.
The magnetic five-brane contributes a source term to the Bianchi identity for the three-form $H$ with
${dH=trR\wedge R-TrF\wedge F-n_5\sum \delta_5^{(4)},}$ where the sum is taken over all five-branes
and $\delta_5^{(4)}$ is a current that integrates to one in the direction 
transverse to a single five-brane whose class is denoted by $[W]$, an element of $H_2(X,{\bf Z})$. Integration over a four cycle in $X$ gives ${{[W]=c_2(TX)-c_2(V).}}$
Now if $V$ has a non-vanishing first Chern class we expect this relation to be modified such that
\eqn\ano{[W]=c_2(TX)+ch_2(V),} 
which is an integer class if $c_1(V)=0 \ ({\rm mod}\ 2)$ or equivalently if $\beta=c_1(B) \ ({\rm mod}\ 2)$.
Note that the latter condition is automatically given for $U(n)$ bundles with $n$ odd as a consistency
condition for $c_1(V)$ and $c_1(L)$. For $n$ even we have to impose $\beta=c_1(B) \ ({\rm mod}\ 2)$ as an additional constraint.

Supersymmetry and the condition to get positive magnetic charges which are carried by the five-brane, 
constrain $[W]$ to be represented by an effective holomorphic curve in $X$. Following the conventions of \DLO, we write the five-brane class as ${W=W_B+a_F F}$
and using \ano\ we find 
\eqn\fivcl{\eqalign{W_B&=\sigma\pi^*(12c_1(B)-\eta)\cr
                                  a_f&=c_2(B)+11c_1(B)^2-\omega+{1\over 8}nc_1(B)^2+{1\over {2n}}\eta_E(\eta_E-nc_1(B))}}
As in the $SU(n)$ case, $W$ is an effective class in $X$ if and only if $W_B$ is effective and $a_F\geq 0$ (this was originally shown in \DLO\ for $SU(n)$ vector bundles). 

\bigskip\noindent{\it An Outlook to F-Theory}\smallskip

In \FMW\ it has been shown that the number of five-branes $a_f$ which are required for the anomaly
cancellation in case $V=E_8\times E_8$ agrees with the number of three-branes $N$ required for anomaly cancellation in $F$-theory on a Calabi-Yau fourfold $Y$ given by ${\chi(Y)\over 24}$. 
If the structure group $G$ of $V$ being contained in $E_8$ then the observed physical gauge group in four-dimensions corresponds to the commutant $\tilde{H}$ of $G$ in $E_8$, typically being of ADE-type. This leads to a generalized physical set-up: The heterotic string compactified on $X$ and a pair of $G$-bundles is dual to $F$-theory compactified on a Calabi-Yau fourfold $Y$ with section and a section $\theta$ of ADE singularities. The generalized set-up has been analyzed in \refs{\FMW} and \refs{\BJPS\AC\ACu-\ACC}. Now if we consider for instance an $U(n)\times E_8$ vector bundle on $X$ we expect a section of singularities in $Y$ corresponding to the commutant of $U(n)$ in the $E_8$ gauge group and thus a modified relation ${\chi(Y)\over24}=a_{E_8}+a_f$ where $a_{E_8}$ corresponds to
the number of five-branes associated to the $E_8$ vector bundle and $a_f$ as given in \fivcl. 

\subsec{The Net-Generation Number}

The net-number of generations $N_{gen}$ in a given heterotic string compactification on a Calabi-Yau
threefold is determined by the index \inds\ if the vector bundle $V$ has non-zero first Chern class. If we insert the expressions for $c_2(X)$ and $ch_i(V)$ we find the net-generation number 
\eqn\indh{N_{gen}=-\gamma C \sigma+{1\over 2}\pi^*((\eta-nc_1)(c_1-\beta))\sigma}
For $U(n)$ vector bundles with $n$ even, we found the consistency condition $\lambda=m$ and $\eta=\beta \ ({\rm mod}\ 2)$ or $\lambda=m+{1\over 2}$ and $\beta=nc_1(B) \ ({mod}\ 2)$ which
guarantees that $c_1(V)$ and $c_1(L)$ are integer classes. In the first case we can take $\lambda=0$
which implied for $SU(n)$ vector bundles the vanishing of $N_{gen}$. However, although the ``gamma''-class vanishes for $U(n)$ bundles, we still have $N_{gen}$ non-zero.

\bigskip\noindent{\it The Enriques Surface Revisited}\smallskip

In case of $SU(n)$ vector bundles on elliptically fibered Calabi-Yau threefolds it has been argued \DLO\
that the Enriques surfaces will never satisfy the effectivity condition for $W_B$ and at the same time 
lead to a realistic net-number of generations. The argumentation is as follows: $W_B$ has to be effective
but $K_B^{\otimes 12}={\cal O}_B$ because 12 is even thus $W_B=-\pi^*\eta \sigma$; since $\eta$ is an effective class it follows $W_B$ can be effective only if $\eta$ is trivial. Now the net-generation number $N_{gen}$ for $SU(n)$ vector bundles is given by \chsdt\ which vanishes for $\eta$ being trivial. 

For $U(n)$ vector bundles the situation does not change which can be seen as follows. We still have to require that $\eta$ is trivial to guarantee the effectivity of $W_B$, this reduces the net-generation number to $N_{gen}={1\over 2}n\pi^*c_1(B)\pi^*\beta\sigma$ which is zero since $N_{gen}$ is a number which can be multiplied by any even number.  

\bigskip\bigskip\noindent
{\bf\centerline{Acknowledgements}}\smallskip
We would like to thank G. Curio, G. Hein, H. Kurke, D. Ploog and E. Sharpe for discussions. 
\noindent
B. A. is supported by DFG Schwerpunktprogramm (1096) ``String Theory im Kontext
von Teilchenphysik, Quantenfeldtheorie, Quantengravitation, Kosmologie und Mathematik''.
\noindent
D. H. R. is supported by DGI research project BFM2003-00097 ``Transformadas Geom\'etricas Integrales y 
Aplicaciones'' and by JCYL research project  SA114/04 ``Aplicaciones de los functores integrales a la Geometr\'{\i}a  y a la F\'{\i}sica''.

\appendix{A}{Notes on the Hermite-Einstein Equations}

In this appendix we recall for the reader's convenience some well-known facts about the Donaldson-Uhlenbeck-Yau and Hermite-Einstein equations and the Hitchin-Kobahashi correspondence. 

Let $X$ be a complex $n$-dimensional K\"ahler manifold with K\"ahler form $J$. We denote by $\Omega^p$ the bundle of complex differential $p$-forms on $X$.

Holomorphic vector bundles can be characterized in terms of connections (see for instance \Kob). We now that if $E$ is a holomorphic vector bundle, then there exist a connection $A\colon E\to E\otimes \Omega^1$  whose $(0,1)$ component $A^{0,1}\colon E \to E\otimes \Omega^{0,1}$ equals the natural $\bar\delta_E$ operator whose kernel is formed by the homolorphic sections of $E$.  The $(0,2)$-component of the curvature $F_A=A\circ A \colon E \to E\otimes \Omega^2$ vanishes. Conversely, whenever we have a connection $A$ on a smooth vector bundle $E$ with $F_A^{0,2}=0$, then there exists a unique holomorphic structure on $E$ such that $A^{0,1}=\bar\delta_E$.

The Chern classes of $E$ can be computed in terms of any complex connection. In particular we have
\eqn\trace{c_1(E)={i\over {2\pi}}Tr (F_A)\,.}

We now consider  hermitian bundles $(E,h)$, that is smooth complex vector bundles $E$ on $X$ endowed with a hermitian metric $h$.  In this case a connection $A$ on $E$ is hermitian if it derives the hermitian metric, that is
$$
d (h(e,e')= h( A(e),e')+ h(e, A(e'))
$$
where $e$, $e'$ are local sections of $E$.  

For hermitian bundles we can refine the above discussion: A holomorphic hermitian bundle has a unique hermitian connection $A$ such that $A^{0,1}=\bar\delta_E$. Moreover the curvature $F_A$ is of type ${1,1}$. We call $S$ the associated hermitian connection to the hermitian holomorphic bundle $(E,h)$.

Conversely, we have the following result: If $(E,h)$ is a hermitian bundle and $A$ is a hermitian connection on  $E$ whose curvature is of type $(1,1)$, then there exists a unique holomorphic structure in $E$ such that $A$ is the associated hermitian connection to the hermitian holomorphic bundle $(E,h)$.

Let $(E,h)$ be a hermitian holomorphic vector bundle and $A$ the associated hermitian connection.  Then $(E,h)$ is said to be a {\sl Hermite-Einstein\/} bundle if the curvature $F_A$ satisfies the {\sl Hermite-Einstein\/} equations
\eqn\ghe{F_A\wedge J^{n-1}=c\cdot I_F\cdot J^n\,,}
where $c$ is a complex constant and $I_E$ is the identity operator on $E$. When the constant $c$ is zero, we get the {\sl Donaldson-Uhlenbeck-Yau\/} equations:
\eqn\gduy{F_A\wedge J^{n-1}=0\,.}
Both Hermite-Enstein and Donaldson-Uhlenbeck-Yau equations can be taught as equations on the hermitian metric $h$ (because the connection $A$ is completely determined by $h$ and the holomorphic structure).

If we take the trace
on both sides of \ghe\ we find
\eqn\yhe{\eqalign{Tr(F_A)\wedge J^{n-1}&=c\cdot Tr(I_E)\cdot J^n\cr
&=c\cdot r\cdot J^n.}}
Now using \trace\ 
we get 
\eqn\inte{\int_X c_1(E)\wedge J^{n-1}={{i\cdot c\cdot r}\over {2\pi}}\int_X J^n}
which determines the constant $c$ (compare \Kob, eqn (2.7)).
With \inte\ we can now write \ghe\ as
\eqn\rec{F_A\wedge J^{n-1}={{2 \pi\cdot c_1(E)\wedge J^{n-1}}\over i\cdot rk(E)}I_E={{2\pi}\over i} \mu(E)J^n I_E}
making apparent that for zero slope $\mu(E)=0$ the Hermite-Enstein equation reduces to the Donaldson-Uhlenbeck-Yau equation
from above.

The Hermite-Einstein conditions for holomorphic vector bundles are equivalent to slope (poly)stability, and this equivalence is known as the Kobayashi-Hitchin correspondence. The precise statement is as follows  (cf. for instance \LuTe):

Let $(E,h)$ be a holomorphic  Hermite-Einstein vector bundle. Then $E$ is $\mu$-polystable (that is, is a direct sum of semistable vector bundles with the same slope). Moreover, if it is irreducible (in the sense that it cannot be represented as a sum of Hermite-Einstein bundles), then $E$ is $\mu$-stable. 

The converse is more subtle: If $E$ is a holomorphic $\mu$-stable vector bundle on $X$, then there exists a hermitian metric $h$ on $E$ such that $(E,h)$ is Hermite-Einstein, that is, $h$ satisfies the Hermite-Einstein equations \ghe.

\appendix{B}{Characteristic Classes of the Fourier-Mukai Transform}

In this appendix se recall from  \OUR\ the formulae giving the topological invariants of both the FM transform and the inverse FM transform of a general complex $\cal G$ is the derived category. They are obtain of course by simply applying GRR for the projection $\pi_2$  to get 
\eqn\grrfp{ch(\Phi({\cal G}))=\pi_{2*}[\pi^*_1({\cal G})\cdot ch({\cal P}) \cdot
Td(T_{X/B})]}
where $Td(T_{X/B})=1-{1\over 2}c_1+{1 \over 12}(13c_1^2+12\sigma
c_1)- {1\over 2}\sigma c_1^2$ (with  $c_1=\pi^*c_1(B)$)
is the Todd class of the relative tangent bundle $T_{X/B}=T_X/\pi^*T_B$.

To compute \grrfp\ we use that the Poincar\'e sheaf is ${\cal P}={\cal O}(\Delta)\otimes {\cal O}(- \pi_1^*\sigma)\otimes {\cal
O}(-\pi_2^*\sigma)\otimes q^*K_B^{-1}$ where ${\cal
O}(\Delta)$ is the dual of the ideal sheaf ${\cal I}$  of the diagonal.  Note first that $ch({\cal I})=1-ch(\delta_*{\cal O}_X)$
with the diagonal immersion $\delta$. Riemann-Roch gives
\eqn\srr{ch(\delta_*{\cal O}_X)Td(X\times_B X)
=\delta_*(ch({\cal O}_X)Td(X))}
where one has the expressions for $Td(X)$ and $Td(X\times_B X)$ given by
\eqn\tdx{\eqalign{Td(X)&= 1+{1\over 12}(c_2+11c_1^2+12\sigma c_1)\cr
        Td(X\times_B X)&=\pi_2^*Td(X)\pi_1^*Td(T_{X/B})}}
The Chern character of the ideal sheaf is then given by
(with the diagonal class $\Delta=\delta_*(1)$)
\eqn\chois{ch({\cal I})=1-\Delta-{1\over 2}\Delta\cdot \pi_2^*c_1+\Delta\cdot \pi_2^*(
\sigma\cdot c_1)+{5\over 6}\Delta\cdot \pi_2^*(c_1^2)+{1\over
2}
\Delta\cdot \pi_2^*(\sigma c_1^2)}
and one can compute $ch({\cal P})$ form that expression. 

Assume now that the Chern characters of the object $E$ of the derived category are of the type
\eqn\chV{ch_0(E)=n_E,\
ch_1(E)=x_E\sigma+\pi^*S_E,\
ch_2(E)=\sigma \pi^*\eta_E+a_E F,\
ch_3(E)=s_E}
($\eta_E, S_E\in H^2(B)$). We note that all the sheaves we are considering (including $V$, $i_*L$ and $W$) have Chern characters of that type. 

It follows from \OUR\ that the Chern characters of the
FM of $E$ and of the inverse FM of
$E$ are
\eqn\oldchVS{\eqalign{ch_0(\Phi(E))&=x_E\cr
ch_1(\Phi(E))&=-n_E\sigma+\pi^*\eta_E-{1\over 2}x_E c_1\cr
ch_2(\Phi(E))&=({1\over 2}n_E c_1-\pi^*S_E)\sigma+(s_E -{1\over 2}\pi^*\eta_E c_1\sigma
+{1\over12}x_E c_1^2\sigma)F\cr
ch_3(\Phi(E))&=-{1\over 6}n_E \sigma c_1^2-a_E +{1\over 2}\sigma c_1
\pi^*S_E}}
and
\eqn\oldchVSS{\eqalign{ch_0(\hat\Phi(E))&=x_E \cr
ch_1(\hat\Phi(E))&=-n_E \sigma+\pi^*\eta_E+{1\over 2}x_E c_1\cr
ch_2(\hat\Phi(E))&=(-{1\over 2}n_E c_1-\pi^*S_E)\sigma+(s_E +{1\over 2}\pi^*\eta_E c_1\sigma
+{1\over12}x_E c_1^2\sigma)F\cr
ch_3(\hat\Phi(E))&=-{1\over 6}n_E \sigma c_1^2-a_E-{1\over 2}\sigma c_1
\pi^*S_E+x_E\sigma c_1^2}}

\appendix{C}{Alternative Derivation of $N_{gen}$ for $SU(n)$ Bundles}

The aim of this appendix is to give an concrete application of the inverse Fourier-Mukai transformation.
We give an alternative derivation of the net-generation number $N_{gen}$ to the derivation given in
\refs{\C,\diac} which adopted the bundle construction of \FMW, i.e.,
consider bundles as given by \vt\ with $c_1(V)=0$. It has been shown in \refs{\C,\diac} that $c_3(V)$ and therefore $N_{gen}$ can be computed from an Riemann-Roch index computation on the matter curve $S=C\cup \sigma$. 
The index reduction was derived using Leray spectral sequence technics \diac. We will now give a shorted derivation of their result. For this we start from \vo\ instead of starting from \vt.  We start with the vector bundle $W$ on $X$ given by \vo\ and 
require $c_1(W)=0$. A similar computation as in section 3.3 gives
\eqn\neco{c_1(L\vert_{S})={1\over 2}(C^2\sigma+C\sigma^2)+\gamma\vert_{S}.}
Now we can apply the Grothendieck-Riemann-Roch theorem and use the expressions for the characteristic classes given in appendix{B} and compute ${1\over 2}c_3(V)$
\eqn\twe{{1\over 2}c_3(V)= {1\over 2}(C^2\sigma+\sigma^2C)-c_1(L\vert_{S})}
Further we have 
\eqn\fourr{2g(S)-2=C^2\sigma+C\sigma^2}
so that
\eqn\thr{{1\over 2}c_3(V)=-(c_1(L\vert_{S})-g(S)+1).}
This shows that the right hand side of \twe\ can be understood 
as a Riemann-Roch index computation on $S$ which is
\eqn\rros{\eqalign{\chi(S,L\vert_{S})&=h^0(S,L\vert_{S})-h^1(S,L\vert_{S})\cr
&=\int_{S}ch(L\vert_{S})Td(S)\cr
&=c_1(L\vert_{S})-g(S)+1}}
thus the standard index computation evaluating $c_3(V)$ reduces to an index
computation on the matter curve $S$
\eqn\stin{\eqalign{{1\over 2}c_3(V)&=h^2(X,V)-h^1(X,V)\cr
&=h^1(S,L\vert_{S})-h^0(S,L\vert_{S})}}
which is in agreement with \diac.

\listrefs
\bye